# A Pilot Study of Sidewalk Equity in Seattle Using Crowdsourced Sidewalk Assessment Data

Chu Li, Allen School of Computer Science, University of Washington

Lisa Orii, Allen School of Computer Science, University of Washington

Michael Saugstad, Allen School of Computer Science, University of Washington

Stephen J. Mooney, Epidemiology, University of Washington

Yochai Eisenberg, Disability and Human Development Department, University of Illinois at Chicago

Delphine Labbé, Disability and Human Development Department, University of Illinois at Chicago

Joy Hammel, Disability and Human Development Department, University of Illinois at Chicago

Jon E. Froehlich, Allen School of Computer Science, University of Washington

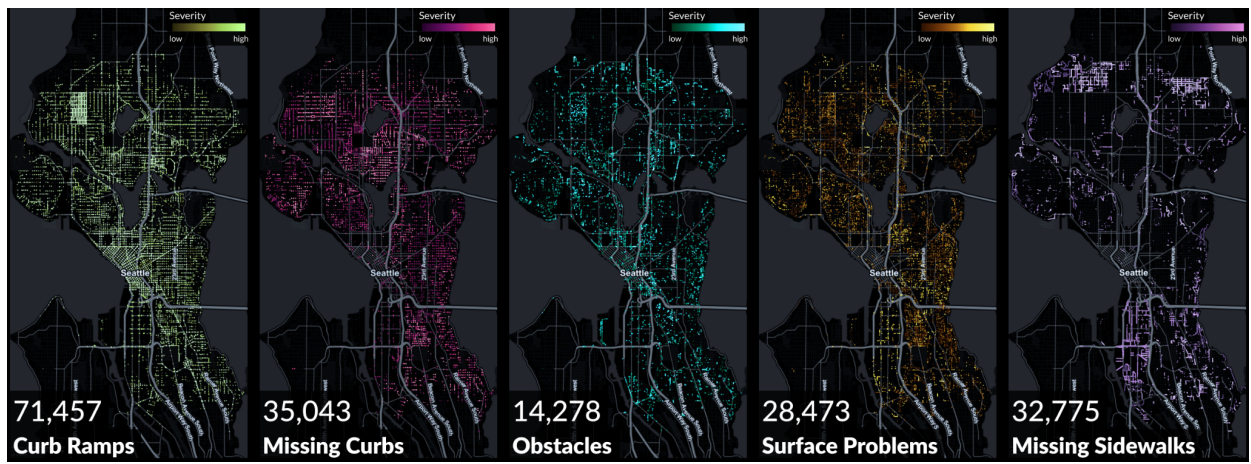

**Figure 1.** A trellis view of the 182,026 geo-located sidewalk condition labels collected in Seattle via Project Sidewalk. With Project Sidewalk, online users virtually investigate sidewalks using Google Street View imagery and mark and assess curb ramps (green), missing curb ramps (pink), obstacles (blue), surface problems (orange) and missing sidewalks (purple). In this workshop paper, we provide a preliminary study examining these crowdsourced sidewalk quality ratings and its relationship to socioeconomic and demographic data.

# Abstract

We examine the potential of using large-scale open crowdsourced sidewalk data from Project Sidewalk to study the distribution and condition of sidewalks in Seattle, WA. While potentially noisier than professionally gathered sidewalk datasets, crowdsourced data enables large, cross-regional studies that would be otherwise expensive and difficult to manage. As an initial case study, we examine spatial patterns of sidewalk quality in Seattle and their relationship to racial diversity, income level, built density, and transit modes. We close with a reflection on our approach, key limitations, and opportunities for future work.

## Introduction

Safe, accessible, and well-maintained sidewalks can enhance public health, promote social interactions, support mobility and independence for people with mobility disabilities and older adults [3,10,30]. Given these benefits, there is growing interest in potential sidewalk-related disparities [17,23,28]—the location, connectivity, and condition of sidewalks—and underlying socio-demographic patterns.

Our research on sidewalk equity is based on data from Project Sidewalk, an open-source crowdsourcing platform that enables online users to remotely label sidewalk conditions and identify accessibility problems using streetscape imagery. As Project Sidewalk uses a combination of automatic and manual techniques for data collection and validation, it is more reliable and accurate than purely automatic object detection algorithms [15] and is more efficient than labor-intensive in situ data collection methods that rely on physical surveying [17].

As early work, our research questions are exploratory:

1. How might we use crowdsourced sidewalk assessment data to examine sidewalk condition patterns in a city?
2. How do sidewalk quality scores relate to neighborhood socioeconomic characteristics?

To address these questions, we present a pilot study of sidewalk equity in Seattle, a seaport city in the US's Pacific Northwest with a metro population of 4 million. We examine spatial patterns of sidewalk quality based on crowdsourced assessments and their relationship to neighborhood characteristics such as racial diversity, income level, built density, education level, and modes of transportation. Within the 363 census block groups studied, we found that sidewalk quality was positively correlated with white population percentage, population density, housing density, and the percentage of people who walk or take public transit to work. Such distinct patterns may be explained by Seattle's urban sprawl development and more recent planning initiatives.

## Sidewalk Equity

Because *equity* has varied definitions and interpretations, we reflect on its use in urban science [11,20,22]. Some studies have defined equity as uniform spatial distribution in a geographical region [4,27], also known as **horizontal equity** or equality. This approach is simple to implement but does not account for variations in spatial distribution that might lead to more social value resulting from sidewalk quality. More recently, urban planners have shifted towards **vertical equity,** wherein disadvantaged groups are disproportionately provided with benefits in order to offset disadvantage [11,19,22]. Metger defines "equity planning" as the responsibility that planners should "*influence opinion, mobilize underrepresented constituencies, and advance and perhaps* *implement policies and programs that redistribute public and private resources to the poor and working class*" [21].

In the space of sidewalk and social equity, studies have been centered around two major topics: *walkability* and *sidewalk conditions*. ***Walkability*** models such as Walk Score® measure the proximity and density of walkable destinations. Studies suggest that people with lower socioeconomic status are likely to live in dense urban areas with higher walkability [8], which may be explained by low car ownership and higher reliance on public transportation [8,24]. Of note, Walk Score® does not account for accessibility issues people with mobility disabilities and older adults may encounter on a daily basis. Meanwhile, *sidewalk conditions* include metrics such as sidewalk presence, width, surface condition, easement, and accessibility features [12,23] . Researchers have found that neighborhoods with lower socioeconomic status have worse sidewalk conditions [17,23]. This may be explained by residential sorting [25], where underprivileged residents systematically relocate to more affordable neighborhoods with poor sidewalk quality. Current studies regarding sidewalk quality are mostly limited to specific regions or scales [1,2,9,17,23] due to data availability challenges [13,15].

In our work, we focus on sidewalk quality, particularly accessibility-related features such as the presence and quality of curb ramps, surface conditions, and obstructions as well as new analytic opportunities presented by crowdsourcing assessments.

## Case Study Seattle

As an initial case study, we examine sidewalk equity in Seattle. In our analysis, neighborhoods refer to geographic regions defined by US census block groups—statistical divisions of census tracts generally defined to contain between 600 and 3,000 people. Block group was chosen as the analysis unit because it best approximates the size of an urban neighborhood and because it is the most detailed geography unit available for socio-economic surveys like the American Community Survey (ACS).

### Datasets

We rely on four primary datasets: Project Sidewalk labels, sidewalk and building block group GIS data from the City of Seattle, and the ACS 2019 (5-year estimates).

**Project Sidewalk Dataset.** Project Sidewalk is an open-source crowdsourcing platform where online users remotely label sidewalk features in Google Street View (GSV), including *curb ramps, missing curb ramps, obstacles, surface problems,* and *missing sidewalks.* For each label, users can provide a severity rating on a 5-point scale (5 is most severe and represents a non-passable barrier for wheelchair users). We used data remotely collected from April 2019 to August 2022: 7,179 users virtually audited 1,196 miles of Seattle's streets (covering 93.8% of the street segments) providing 211,350 geo-located sidewalk accessibility labels. Of these labels, 53,791 were removed either due to being from tutorial sessions (10.1%) or inferred as "low quality" by Project Sidewalk's built-in quality inference algorithm (15.3%). Because multiple users can label the same sidewalk feature, Project Sidewalk clusters proximal labels of the same type together—see Clustering in [25]. The remaining 157,559 labels were grouped into 95,879 clusters. 93,201 clusters were used in the study after removing data without severity ratings.

**Seattle Sidewalk GIS data.** To associate Project Sidewalk's point assessments with the underlying sidewalk network, we use sidewalk geometry data from Seattle's *Open Data Portal* [5] (46,193 sidewalk segment records).

**Seattle Block Group GIS.** The Seattle Block Group geographic data was obtained from Seattle's *Open Data Portal* [6]. In our study, we used 363 out of 482 block groups in Seattle where we have Project Sidewalk and Sidewalk GIS data.

**Socio-demographic data.** To explore sidewalk conditions with socio-economic factors on a neighborhood level, we obtained data from the 2019 American Community Survey (ACS) 5-year Estimates—see Table 1.

Table 1. American Community Survey (ACS) census block group data of sociodemographic variables used in the study. To account for the varying population in each block group, some indicators were calculated as proportions. (*e.g.*,instead of using the absolute number of people with bachelor's degrees, we used the proportion of bachelor's degrees). Racial diversity was calculated using Simpson's diversity index [26], which measures the probability that any two people chosen from the same block group at random, would be of different races.

| Categories | Factors | Details | Metric |
|---|---|---|---|
| Demographic | Density | Population Density Per Sq. Mile | Count |
| | Race | White, Black or African American; Asian; American Indian and Alaska Native; Native Hawaiian and Other Pacific Islander; Some Other Race Alone; Two or More Races; Racial Diversity | Proportions |
| | Household | Family Households; Nonfamily Households | Proportions |
| Housing | Housing value | Median House Value; Median Gross Rent | Count |
| | Units in structure | Housing Units: 1(Detached); 1(Attached); 2; 3 or 4; 5-9; 10-19; 20-49; 50 or more; Housing Age | Proportions |
| | Tenure | Owner Occupied Housing; Renter Occupied Housing | Proportions |
| Social | Citizenship | Native Born; Foreign Born (naturalized); Foreign Born (not a citizen) | Proportions |
| | Education attainment | Less than High School; High School; Graduate Some College; Bachelor's Degree; Master's Degree; Professional School Degree; Doctorate Degree | Proportions |
| Economic | Income | Median Household Income; Per Capita Income | Count |
| | Employment | Employed; Unemployed | Proportions |
| | Commute | Car, Truck, or Van; Public Transportation; Motorcycle; Bicycle; Walked | Proportions |

## Constructing Sidewalk Accessibility Scores

Creating a representative scoring model for the accessibility and condition for sidewalks is an open research question [13,16,18], yet a reliable and grounded sidewalk scoring method is crucial to the presented work. Prior work using Project Sidewalk data has calculated accessibility scores based on grids [18], streets [14], and neighborhoods [14]. Here, we base our approach on Hara [14], which calculates *Access Scores* at street- and neighborhood-levels, and extend this method by incorporating severity ratings.

**AccessScore: Sidewalk Segment** ($AS_{sidewalk}$) measures the accessibility level of a sidewalk segment with *high* scores corresponding to *higher* accessible sidewalks. To calculate an *AccessScore*, we first assign labels to the closest sidewalk geometry (Fig2a, b) and then group labels by type and severity rating to generate an **accessibility feature vector** ($xa$).

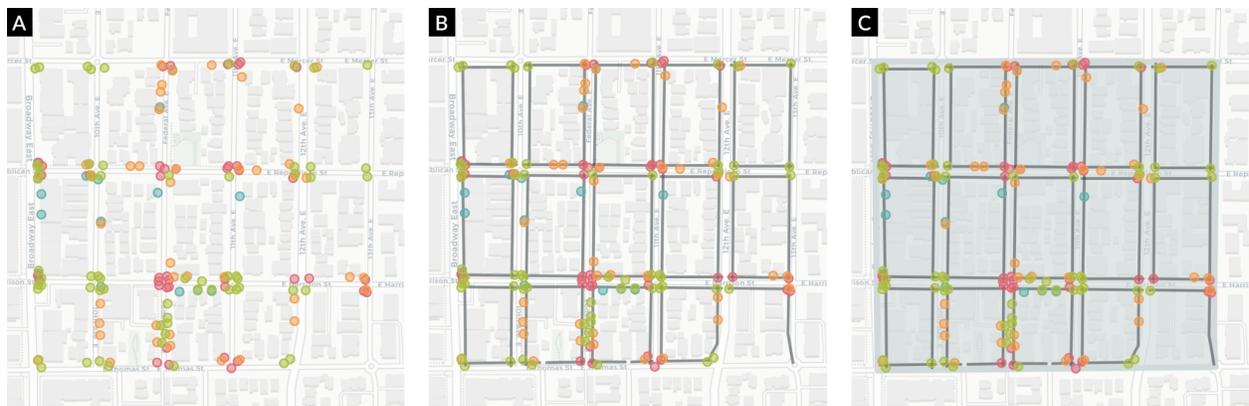

**Figure 2.** Example of joining datasets in a block group. We first assigned Project Sidewalk labels (a) to their closest sidewalk segments (b) by performing a nearest neighbor join using GeoPandas, with a maximum distance of 30 feet. Sidewalk segments were then mapped to their intersecting block group polygon (c). Subsequently, socio-economic data from ACS were linked to the corresponding block group through a geographic identifier.

We then take a dot product of the accessibility feature vector and a **significance vector** ($ws$), a vector representing the importance of each accessibility feature type. Each significance vector element has a value between 0.2−1 converted from severity ratings 1−5. For example, a curb ramp of severity 5 (non-passable for wheelchair users) is discounted as 0.2 of a curb ramp. The significance vector's polarity (+/-) indicates whether it is a positive or negative accessibility feature. Finally, as the range of the dot product could be between (−∞, ∞), we map it to (0, 1) using a sigmoid function. See example in Figure 3.

In summary, $AS_{sidewalk}$ is calculated as : $AS_{sidewalk} = \frac{1}{1 + e^{-(ws \cdot xa)}}$

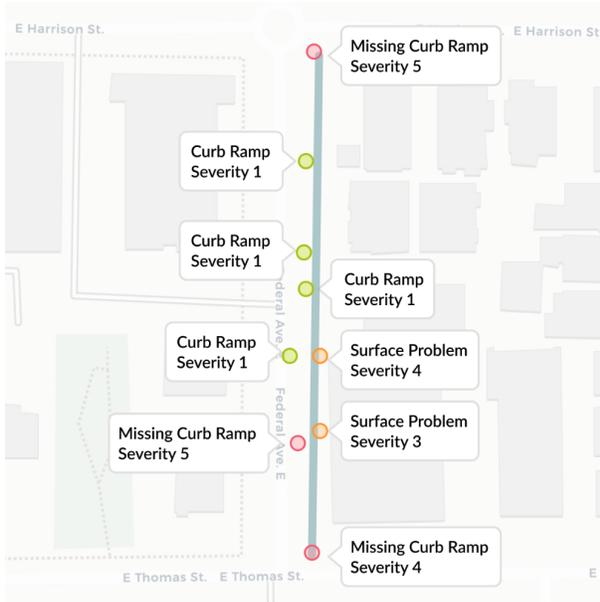

The sidewalk segment shown on the left has:

4* Curb Ramp_Severity 1,
1* Missing Curb Ramp_Severity 4,
2* Missing Curb Ramp_Severity 5,
1* Surface Problem_Severity 3,
1* Surface Problem_Severity 4,

**Accessibility Feature Vector**

$ws$ = (4, 1, 2, 1, 1)

**Significance Vector**

$xa$ = (1.0, -0.8, -1.0, -0.6, -0.8)

**AccessScore: Sidewalk Segment**

$AS_{sidewalk} = \frac{1}{1+e^{-(ws \cdot xa)}}$ = 0.45

**Figure 3.** Example of how AccessScore: Sidewalk Segment is calculated.

**AccessScore: Neighborhood** ($AS_{neighborhood}$) We calculate the *AccessScore* of a neighborhood by totaling the weighted accessibility score of all the sidewalks contained within a given block group's polygon, which is normalized by the total length of sidewalks in that block group:

$$AS_{neighborhood} = \sum_{i=1}^{n} \frac{AS_{sidewalk} \times length_{sidewalk}}{length_{total}}$$

Here, $n$ represents a number of sidewalk segments contained in a given block group polygon.

## Findings

Below, we present an overview of sidewalk accessibility in Seattle as well as our correlation analysis. Our analysis scripts were written in Python with GeoPandas in Jupyter Notebook—see our open source repo here: https://github.com/ProjectSidewalk/sidewalk-equity-study.

## Sidewalk Accessibility Distribution

The $AS_{sidewalk}$ scores ranged from 0.00 to 0.99 (*avg*=0.49) while the $AS_{neighborhood}$ ranged 0.06 to 0.89 (*avg*=0.47)—see distributions in Figure 4.

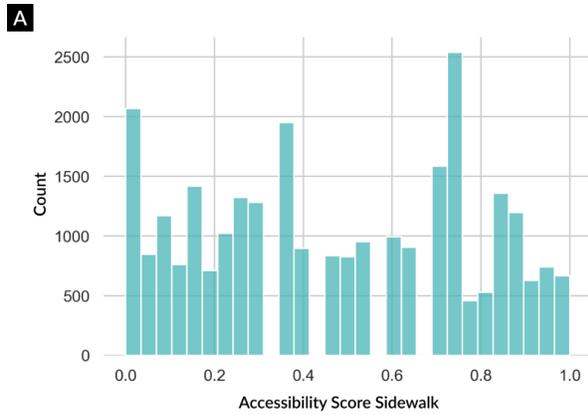
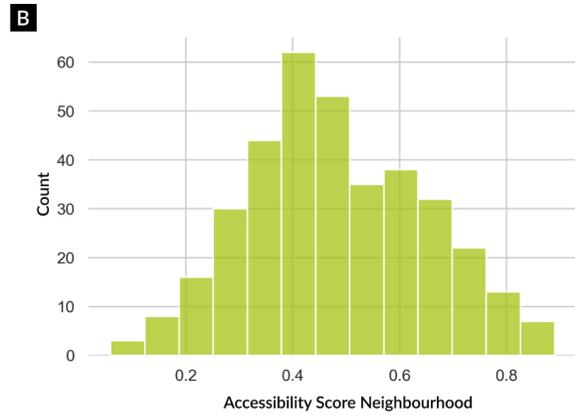

**Figure 4.** Access score distribution per sidewalk segment (a) and access score distribution per neighborhood (b).

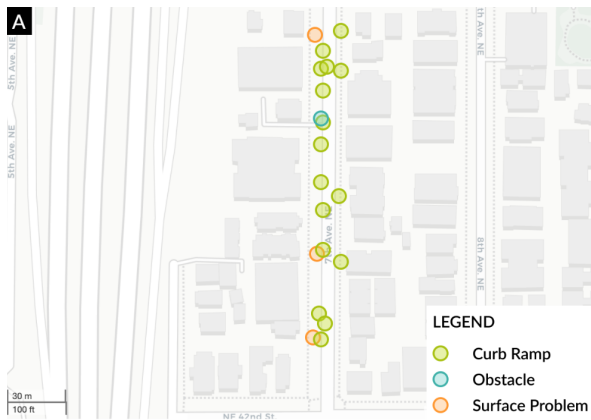
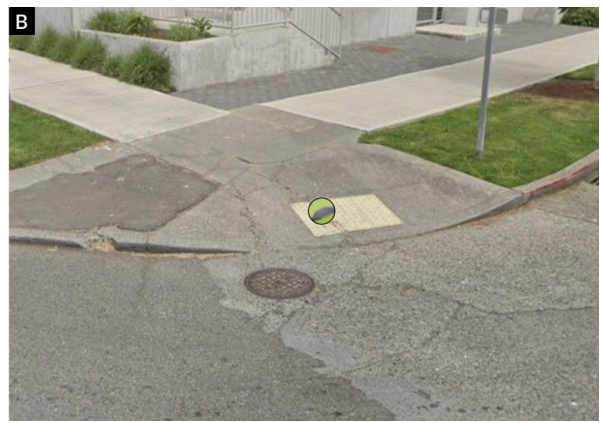

**Figure 5.** (a) A sidewalk segment with a high access score of 0.99 (b) and a curb ramp example from the same sidewalk segment.

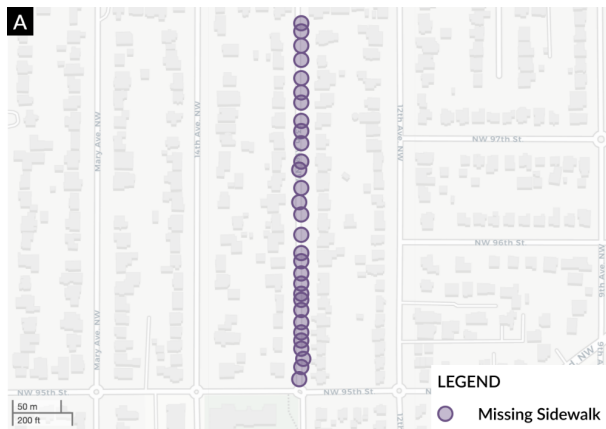
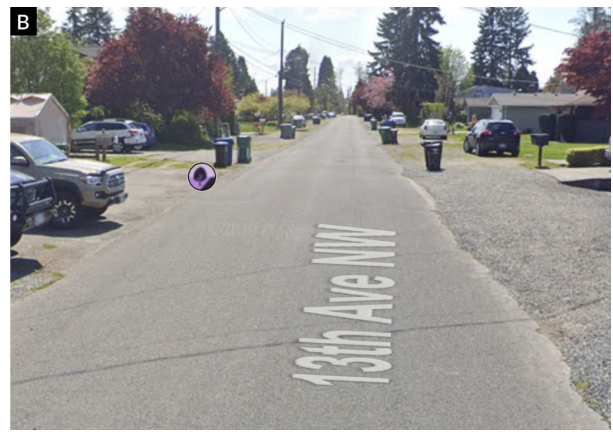

**Figure 6.** (a) A sidewalk segment with a low access score of 0.00 (b) and a missing sidewalk example from the same sidewalk segment.

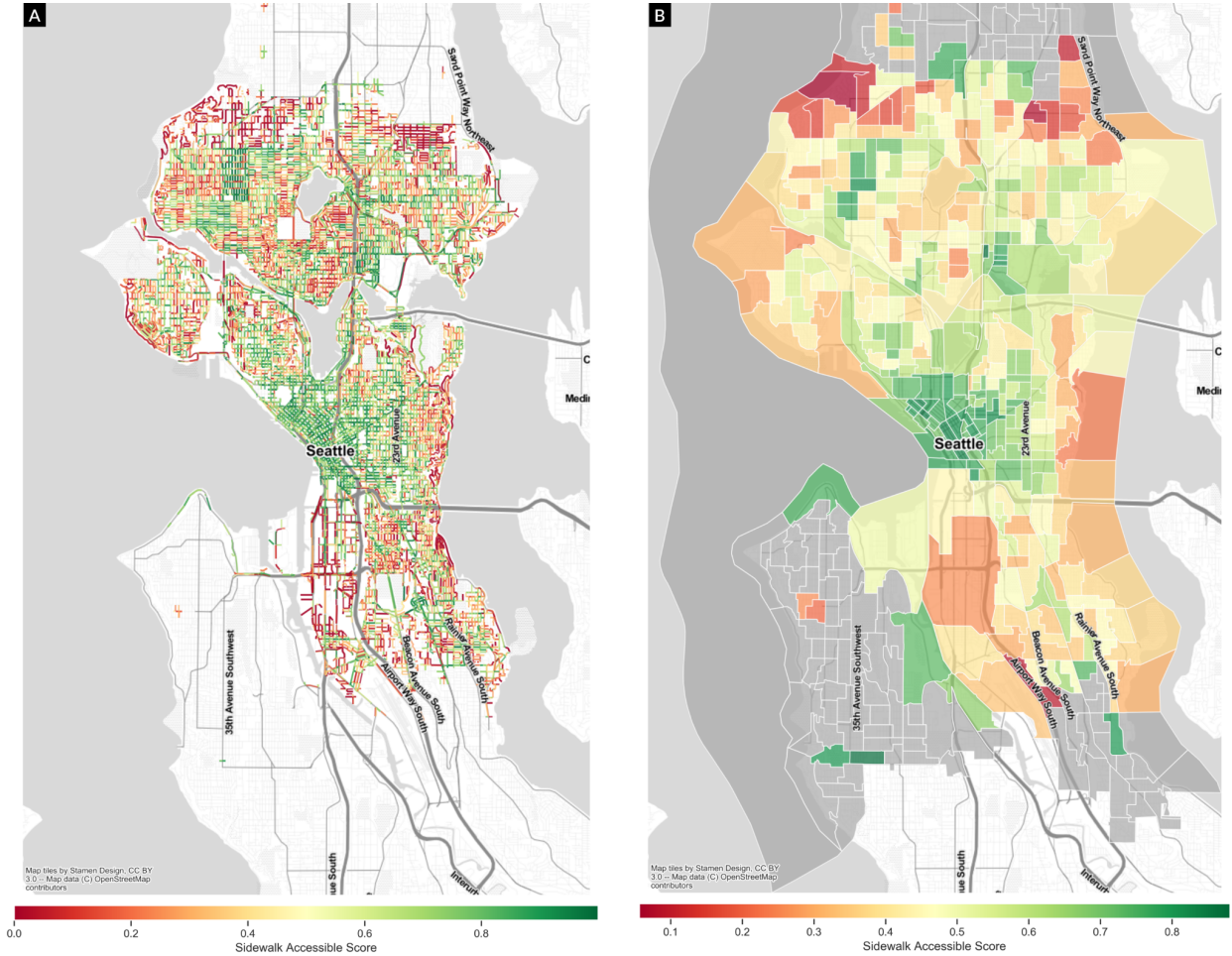

**Figure 7.** (a) *AccessScores* per sidewalk segment and (b) *AccessScores* per neighborhood based on Project Sidewalk data. Higher accessibility scores reflect better accessibility for people who use wheelchairs (in green).

In both mapping access scores per sidewalk and per neighborhood, we observed a radial pattern with a high accessibility score in the urban core and a lower accessibility score towards the periphery of the study area.

## Sidewalk accessibility vs socio-economic factors

Since demographic data were not normally distributed, Spearman's rho was used to analyze correlation between demographic characteristics and sidewalk accessibility. The results are summarized in Table 2. After filtering out the weak correlations (ρ <= ±0.2), we found that sidewalk accessibility is positively correlated with population density; the percentage of non-white population; the percentage of non-U.S. citizens; the percentage of renter occupied housing; the percentage of housing with 10 or more units, the percentage of housing built after 1990, the percentage of people who walked or took public transportation to work. Sidewalk accessibility is inversely correlated with the percentage of white population, racial diversity, the percentage of family households, the average household size, household income (both average and median), the percentage of owner occupied housing, the percentage of single unit detached housing, the percentage of housing built between 1940 and 1959, housing median value, and the percentage of people who drive to work.

**Table 2.** Spearman's rho correlation coefficients of the American Community Survey (ACS) census block group data of sociodemographic variables and sidewalk accessibility score per block group (n = 363). (*p < 0.1, **p < 0.05, ***p <0 .01). Green cells indicate positive correlations, red cells indicate negative. MTW stands for *means of transportation to work*, TTW stands for *travel time to work*.

| Property | rho | p | Property | rho | p | Property | rho | p |
|---|---|---|---|---|---|---|---|---|
| Population Density (Per Sq. Mile) | 0.52 | *** | MTW: Car, Truck, or Van% | 0.50 | *** | Average Gross Rent | -0.13 | ** |
| White Alone% | -0.23 | *** | MTW: Drove Alone% | 0.49 | *** | Owner Occupied% | 0.57 | *** |
| Black or African American Alone% | 0.12 | ** | MTW: Carpooled% | -0.24 | *** | Renter Occupied% | 0.57 | *** |
| American Indian & Alaska Native Alone% | 0.16 | *** | MTW: Public Transportation% | 0.23 | *** | Units in Structure: 1, Detached% | 0.60 | *** |
| Asian Alone% | 0.23 | *** | MTW: Motorcycle% | -0.07 | | Housing Units: 1, Attached% | -0.04 | |
| Pacific Islander Alone% | 0.07 | | MTW: Bicycle% | -0.07 | | Units in Structure: 2% | -0.07 | |
| Some Other Race Alone% | 0.14 | *** | MTW: Walked% | 0.50 | *** | Units in Structure: 3 or 4% | 0.03 | |
| Two or More Race% | -0.01 | | MTW: Other Means% | 0.04 | | Units in Structure: 5 to 9% | 0.07 | |
| Racial Diversity | -0.23 | *** | TTW: Less than 10 Minutes% | 0.13 | ** | Units in Structure: 10 to 19% | 0.20 | *** |
| Citizenship - Native% | -0.08 | | TTW: 10 to 19 Minutes% | 0.21 | *** | Units in Structure: 20 to 49% | 0.43 | *** |
| Foreign Born - Naturalized% | -0.17 | *** | TTW: 20 to 29 Minutes% | 0.07 | | Units in Structure: 50 or More% | 0.51 | *** |
| Foreign Born - Not a Citizen% | 0.23 | *** | TTW: 30 to 39 Minutes% | -0.14 | *** | Housing Units Built 2014 or Later% | 0.35 | *** |
| Family Households% | 0.49 | *** | TTW: 40 to 59 Minutes% | -0.20 | *** | Housing Units Built 2010 to 2013% | 0.27 | *** |
| Average Household Size | 0.42 | *** | TTW: 60 to 89 Minutes% | -0.13 | ** | Housing Units Built 2000 to 2009% | 0.28 | *** |
| Less than High School% | 0.12 | ** | Median Household Income | -0.31 | *** | Housing Units Built 1990 to 1999% | 0.22 | *** |
| High School Graduate% | 0.00 | | Average Household Income | -0.32 | *** | Housing Units Built 1980 to 1989% | 0.07 | |
| Some College% | 0.04 | | Median Family Income | -0.15 | *** | Housing Units Built 1970 to 1979% | -0.02 | |
| Bachelor's Degree% | -0.02 | | Average Family Income | -0.13 | ** | Housing Units Built 1960 to 1969% | -0.10 | * |
| Masters Degree% | -0.01 | | Per Capita Income | -0.06 | | Housing Units Built 1950 to 1959% | -0.33 | *** |
| Professional School Degree% | -0.09 | * | Median Housing Value | -0.21 | *** | Housing Units Built 1940 to 1949% | 0.42 | *** |
| Doctorate Degree% | -0.13 | ** | Median Gross Rent | -0.13 | ** | Housing Units Built 1939 or Earlier% | -0.12 | ** |
| Unemployed% | 0.07 | | Median Gross Rent as a % of Income | 0.16 | *** | | | |

# Discussion

In this paper, we presented a preliminary study of sidewalk equity in Seattle using crowdsource datasets, below we reflect on our key findings and limitations.

## Sidewalk accessibility patterns and socio-economic factors

Our analysis found that neighborhoods with lower sidewalk quality are more affluent, predominantly white, have a higher percentage of families, have lower housing and population density, and use driving as their primary mode of transportation. Neighborhoods of this type are located in the periphery of the study area. In contrast, neighborhoods with higher sidewalk accessibility have higher population and housing density, are more racially diverse, have a higher proportion of immigrants and commute primarily by walking or public transportation. These neighborhoods are located within, or immediately adjacent to, the urban core. While our results differ from some previous studies that reported poor sidewalk quality in underprivileged neighborhoods [17,23], the sidewalk disparity we identified seems to be more closely related to urban sprawl and more recent planning initiatives.

**Urban sprawl development.** Nearly 24 percent of Seattle's streets have missing sidewalks [32], most of which are located in north Seattle, an area which was annexed from King County in the 1950s. As King County lacked Seattle's sidewalk development requirements, the annexed area did not include many sidewalks, and the condition persists today [7,29].

**Recent planning efforts.** The City of Seattle has acknowledged its sidewalk problem and is working towards improving the presence and condition of sidewalks through *Levy to Move Seattle* [33] and the *Sidewalk Repair Program* [34]. Seattle's growth strategy emphasizes that most of the city's growth should occur in urban centers and manufacturing/industrial centers [35], priority investments for sidewalk improvements have also been made to the densest neighborhoods [32]. In 2017, Seattle settled a federal class-action lawsuit to fix or install 22,500 curb ramps over 18 years at an estimated cost of $300m [31]. Crowdsourcing tools, like Project Sidewalk, could help track these changes and ensure accountability.

## Limitations

**Data collection.** Project Sidewalk's labeling data does not cover all sidewalk geometries in Seattle. There are 101 miles of sidewalk that remain un-audited—5.6% of all sidewalks in the city. Our analysis should be repeated once Project Sidewalk data is fully completed.

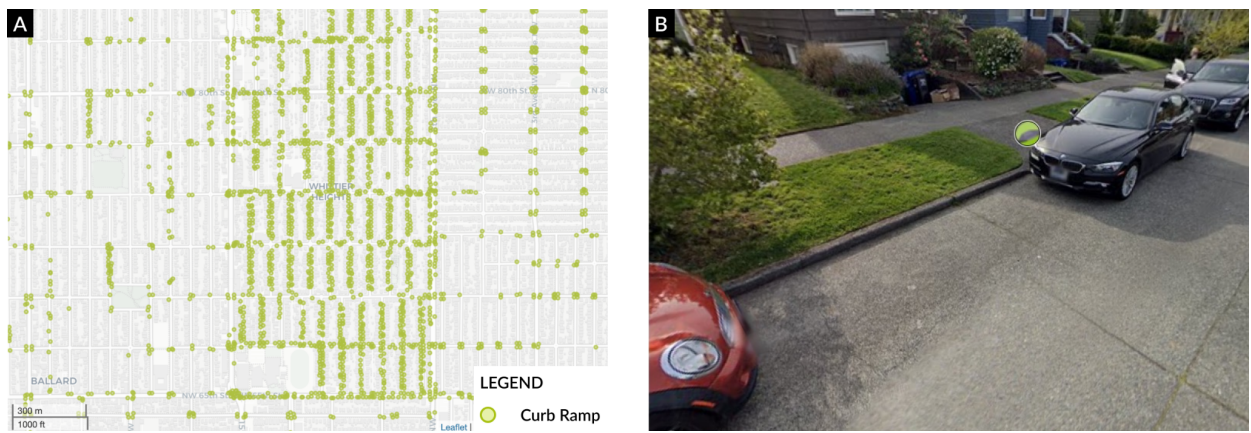

**Figure 8.** As an example of data noise in Project Sidewalk, this user labeled both legitimate curb ramps (correctly) and driveways as curb ramps (incorrectly). While driveways are often used as a last-resort accessible pathway, they are not ADA regulated and should not be labeled as curb ramps. Such false positives and other sources of data noise can impact our analyses and should be explored further.

**Data quality.** All data collection methods have strengths and weaknesses. As one metric of quality, the data used in our study has 150k crowdsourced validations and a 76.1% agreement rate. More work is needed to assess crowdworker reliability and its impact on correlative analyses. See Figure 7 for a particularly extreme example of noisy data.

**Data age.** Currently, Project Sidewalk labels do not include information regarding when the label was created and when the corresponding Google Street View image was captured. Our results may include errors arising from time mismatches. However, as both accessibility and social conditions change slowly, Street View imagery in Seattle is updated frequently, and our labels date from 2019-2022, we do not anticipate substantial errors would arise from this mismatch.

**AccessScore calculations.** Our pilot study employed simple methods that focused only on accessibility features within a given sidewalk segment or neighborhood. A variety of other factors can be considered when computing accessibility scores, including terrain information, temporary accessibility barriers, and Walk Score®.

## Future Work

As early work, this study opens up more questions than it answers: how can we develop algorithmic models to accurately reflect sidewalk accessibility? What spatial scale (*e.g.*, block group, census tract, zip code) best portrays the relationship between sidewalk quality and social equity? To address the needs of different groups and individuals, how can we create an interactive platform that offers customizable analyses and visualizations? How can local governments utilize these studies to prioritize and direct limited planning resources towards ADA renovations? In future studies, we will attempt to address these questions with the hopes of fostering a more livable and equitable urban future.

## Acknowledgments

This work was funded in part by the National Science Foundation under grant SCC-IRG 2125087, the Pacific Northwest Transportation Consortium (PacTrans), and UW CREATE.

# Author Bios

**Chu Li** is an incoming PhD student in computer Science at UW interested in enhancing the design of cityscapes through technology. She has a MS in Information Science from the University of Toronto, a MS in Architecture and Urban Design from Columbia University. She previously worked as an urban designer at Skidmore, Owings & Merrill LLP (SOM), Chicago.

**Lisa Orii** is a PhD student in Computer Science at UW interested in exploring equitable technologies in healthcare settings in low-income and low-resource communities. Lisa is a Funai Foundation Fellow and has two undergraduate degrees from Wellesley College in CS and Philosophy.

**Michael Saugstad** is Project Sidewalk's lead engineer and research scientist at the University of Washington.

**Stephen J. Mooney** is an Assistant Professor in Epidemiology at UW and is interested, broadly, in how cities affect health, and how we can know whether they do. Stephen is one of the co-creators of the Computer-Assisted Neighborhood Visual Assessment System (CANVAS) to measure health-related neighborhood conditions.

**Yochai Eisenberg** is an Assistant Professor in Disability and Human Development at UIC studying how neighborhood environments, local policies, and systems impact health behaviors and outcomes for people with disabilities using big data analytics, policy evaluation, and community engaged research. Yochai developed the Community Health Inclusion Index (CHII), which can be used by communities to increase participation of people with disability in health promoting activities

**Delphine Labbé** is an Assistant Professor in Disability and Human Development at UIC with a focus on promoting full participation and health of people living with disabilities by optimizing their interaction with the social and physical environment.

**Joy Hammel** is a Professor in Disability and Human Development and the Wade Meyer Endowed Chair in Occupational Therapy at UIC. Joy's research and teaching focus on community-based participatory action research related to community living and participation choice, control and societal opportunity or disparities with people who are aging with disabilities and disability & aging communities

**Jon E. Froehlich** is an Associate Professor in Computer Science at UW, the Director of the Makeability Lab, and the Associate Director of the Center for Research and Education on Accessible Technology and Experiences (CREATE). He is interested in applying Human+AI methods to transform how we study and analyze cities and urban accessibility.

## Rationale for Attendance

Our work on sidewalk and equity touches on many of the key questions of this workshop such as: *what is the role of AI in assessing urban accessibility and how can we create effective visualizations to support new policy and urban planning?* In demonstrating the potential of using large-scale, open crowdsourced sidewalk data to study sidewalk conditions and social factors, we hope to scale our research and ultimately be able to examine the *differences in urban accessibility needs across the globe.* Through AccessScore, we attempted to develop algorithmic models to accurately reflect sidewalk accessibility, working towards *creating personalized interactive models of urban accessibility.* Our team members come from diverse backgrounds spanning across fields of computer science, human-computer interaction, accessibility, epidemiology and urban planning. It is our hope to contribute our distinct domain knowledge and perspectives, as well as benefit from the interdisciplinary ideas sparked during this workshop.